\documentclass[slac_one]{revtex4}
\usepackage{graphicx}
\usepackage{wrapfig}
\usepackage{fancyhdr}
\begin{document}

\title{Heavy Quark Electroproduction and the Heavy Quark Contribution 
to the Proton Structure} 

%

\author{K. Kr\"uger (for the H1 and ZEUS Collaborations)}
\affiliation{Kirchhoff-Institut f\"ur Physik, Universit\"at Heidelberg, 
69120 Heidelberg, Germany}

\begin{abstract}
New results on charm and beauty electroproduction from the H1 and ZEUS 
Collaborations at the $ep$ collider HERA are presented. Differential cross
sections are compared to a next-to-leading order QCD calculation. 
The heavy quark contributions, $F_2^{c {\bar c}}$ and $F_2^{b {\bar b}}$,
to the proton structure function $F_2$ are compared using different 
experimental techniques. Also, results are shown on the charm fragmentation 
function, thereby testing the universality of fragmentation.
\end{abstract}

\maketitle

\thispagestyle{fancy}

\section{INTRODUCTION}
At HERA, $27.5\,{\rm GeV}$ electrons (or positrons) were collided with 
$920\,{\rm GeV}$ protons providing a center-of-mass energy of 
$\sqrt{s} = 318\,{\rm GeV}$. The H1 and ZEUS experiments each collected data
corresponding to an integrated luminosity of about $500\,{\rm pb}^{-1}$,
split into two running phases: HERAI and HERAII. The
dominant production process for heavy quarks at HERA is photon gluon fusion,
in which a photon emitted from the electron fuses with a gluon in the proton
producing a quark anti-quark pair. Thus the production of heavy quarks is 
directly sensitive to the gluon density in the proton. In the case of 
electroproduction or deep-inelastic scattering (DIS), the virtuality of the 
exchanged photon is large ($Q^2
\mbox{$\;\raisebox{-1mm}{$\stackrel{\textstyle>}{\sim}$}\;$}2\,{\rm GeV}^2$) and
provides a hard scale for reliable calculations in perturbative Quantum 
Chromodynamics (pQCD). The available pQCD calculations are performed in a 
scheme ("massive") which assumes no heavy quark content of the
proton and are based on collinear or $k_T$ factorisation.

\section{CROSS SECTION MEASUREMENTS}
\subsection{Charm Measurements}
\begin{figure}
\includegraphics[width=0.48\columnwidth]{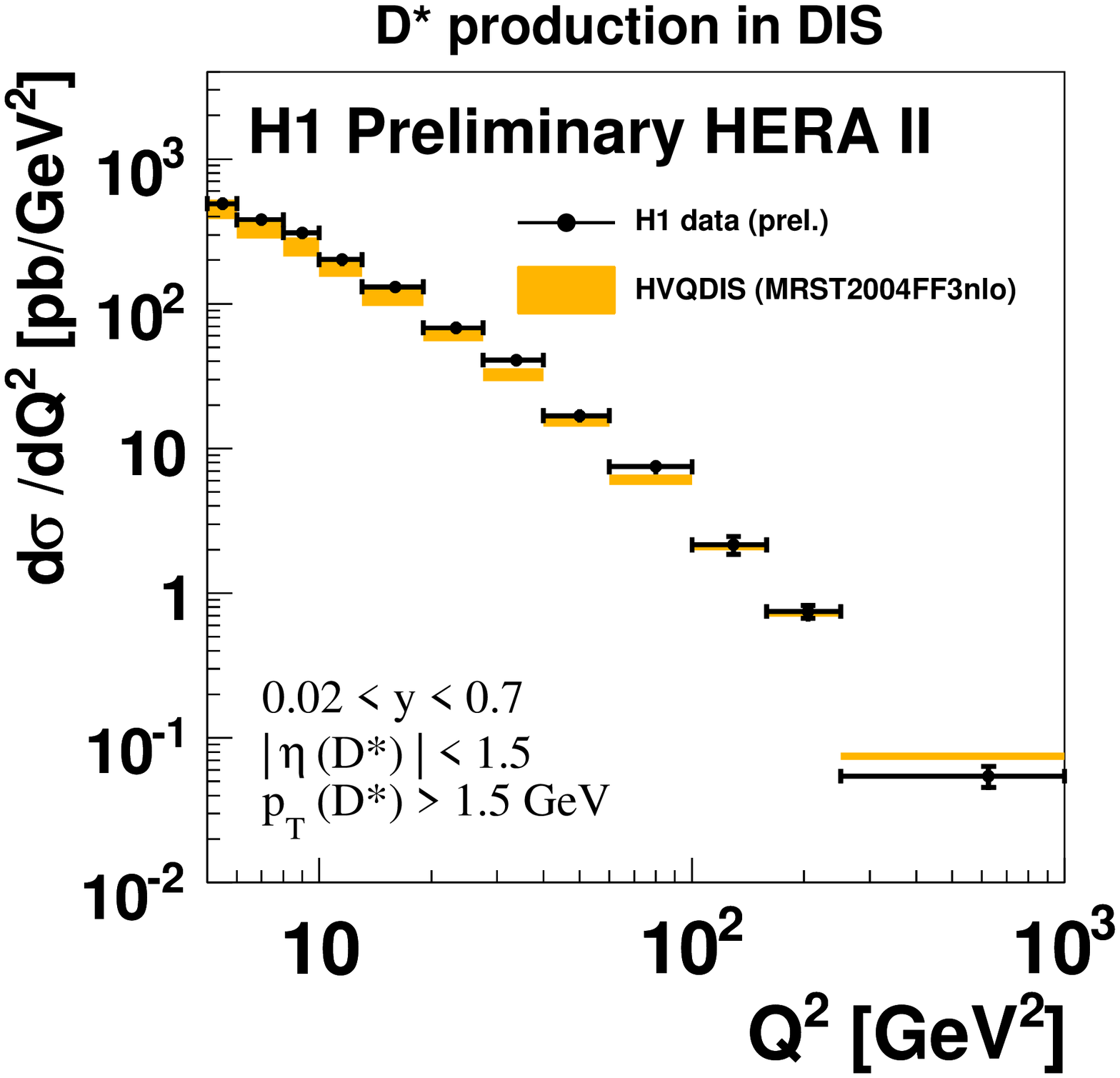} \hfill
\includegraphics[width=0.47\columnwidth]{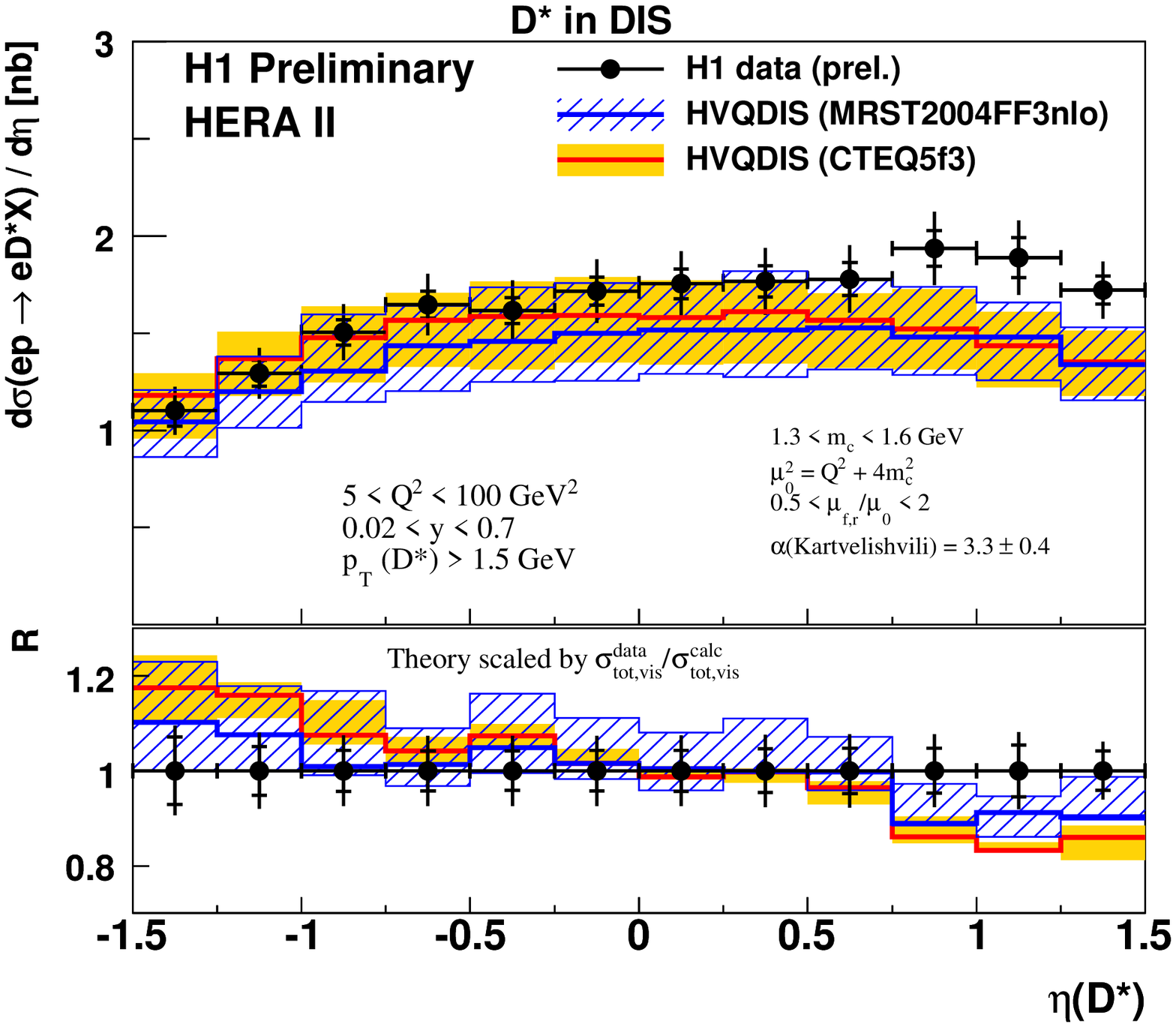}
\caption{$D^{*\pm}$ cross section as a function of (a) the photon virtuality 
$Q^2$ and (b) the pseudorapidity of the $D^{*\pm}$ meson $\eta(D^{*})$ 
compared to the NLO QCD calculation HVQDIS with the proton parton
densities MRST2004FF3nlo~\cite{mrst} and CTEQ5F3~\cite{cteq}.
\label{Fig:dstar_q2}}
\end{figure}

Many experimental results for charm production at HERA use charm mesons
to identify the presence of charm quarks, especially the $D^{*\pm}$ meson
exploiting the well known mass difference method.
The H1 Collaboration presented new preliminary measurements of the 
electroproduction of
$D^{*\pm}$ mesons in two regions of the photon virtuality: 
$5<Q^2<100\,{\rm GeV}^2$~\cite{h1dstarmedium} and 
$100<Q^2<1000\,{\rm GeV}^2$~\cite{h1dstarhigh}. The visible phase space
of the $D^{*\pm}$ meson is restricted to $|\eta(D^{*})|<1.5$ and 
$p_T(D^{*})>1.5\,{\rm GeV}$. The results are
based on the full HERAII statistics corresponding to an integrated luminosity
of about $350\,{\rm pb}^{-1}$. The next-to-leading order (NLO) massive QCD 
calculation HVQDIS~\cite{hvqdis} (using collinear factorisation) agrees with 
the measurement up to the largest $Q^2$ (fig.~\ref{Fig:dstar_q2} (a)) where 
the massive approach is not expected to be appropriate.
The single and double differential cross sections as a function of 
the transverse momentum $p_T$ and the 
pseudorapidity $\eta$ (fig.~\ref{Fig:dstar_q2} (b)) of the $D^{*}$ 
meson are reasonably well described by HVQDIS and show a sensitivity to the 
gluon density in the proton. Unfortunately, the theoretical uncertainties 
due to a variation of the charm mass, the renormalisation and factorisation 
scales and the fragmentation parameter $\alpha$ are larger than the 
experimental uncertainties.   

The ZEUS Collaboration has measured differential cross sections for
$D^{*\pm}$~\cite{zeusdstar}, $D^0$ and $D^{\pm}$~\cite{zeusdnullplus} mesons 
in the range $5<Q^2<1000\,{\rm GeV}^2$ based on part of the HERAII luminosity.
For the $D^0$ and the $D^{\pm}$ analysis the signal-to-background ratio has 
been improved by using lifetime information from the Micro Vertex Detector.
The HVQDIS prediction describes the $Q^2$, $p_T$ and $\eta$ distributions for
all charm mesons reasonably well.

In order to study the charm production process further the angular correlation 
of dijets in events with $D^*$ mesons was measured by the H1 
experiment~\cite{h1dstardijet}. At small azimuthal 
differences between the two jets, where higher order processes are expected to
contribute, the HVQDIS calculation as well as the leading order (LO) 
Monte Carlo program CASCADE~\cite{cascade}, based on $k_T$ factorisation and 
including parton showers, have problems to describe the data.

\subsection{Beauty Measurements}
\begin{wrapfigure}{r}{0.58\columnwidth}
\centerline{\includegraphics[width=0.53\columnwidth]{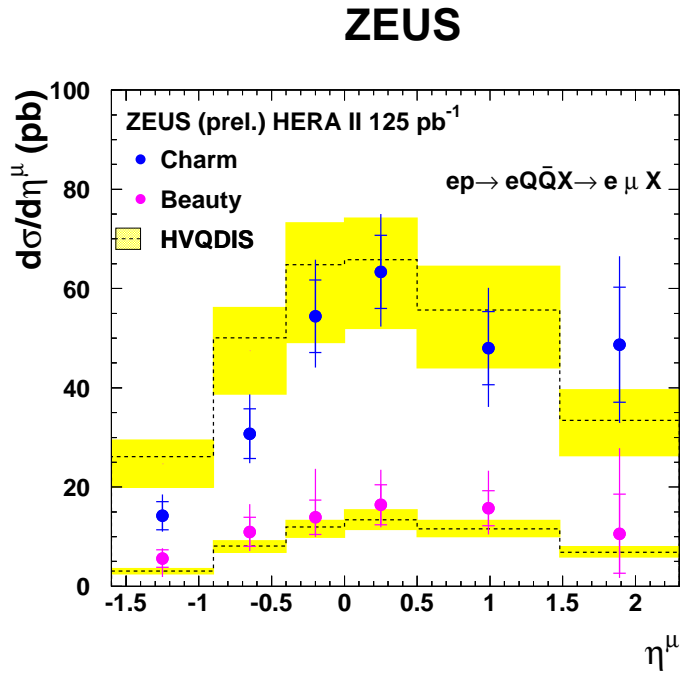}}
\caption{Cross section for the production of a muon and a jet from the decay of 
beauty and charm quarks as a function of the pseudorapidity of the muon
$\eta_\mu$.
\label{Fig:beauty_eta}}
\end{wrapfigure}
Since the beauty cross section at HERA is very small, the reconstruction of
exclusive hadron decays is difficult. Therefore, more efficient tagging methods
which exploit either the long lifetime of $B$ hadrons using precise silicon
vertex detectors (displaced track method) or the large $b$ quark mass using
the transverse momentum of leptons with respect to jets ($p_{T}^{rel}$ method),
are used instead. A new preliminary measurement~\cite{zeusbmujet} combining 
the methods in events with a muon and a jet was presented by the ZEUS 
experiment. The kinematic range was restricted to $Q^2>20\,{\rm GeV}^2$, 
$p_T^\mu > 1.5\,{\rm GeV}$ and $-1.6<\eta^\mu < 2.3$. As the jet was used only 
for tagging purposes, a loose jet
selection of $p_T^{jet}>3\,{\rm GeV}$ and $-3<\eta^{jet} < 3$ was applied.
The cross sections were extracted by a three-dimensional fit to
the impact parameter, the momentum of the muon transverse to the jet 
and the missing transverse momentum parallel to the muon direction. The 
measured differential beauty and charm cross sections are reasonably well 
described by HVQDIS (fig.~\ref{Fig:beauty_eta}).

\section{CONTRIBUTION TO THE PROTON STRUCTURE FUNCTION \boldmath $F_2$}
\begin{figure*}[t]
\includegraphics[width=0.42\columnwidth]{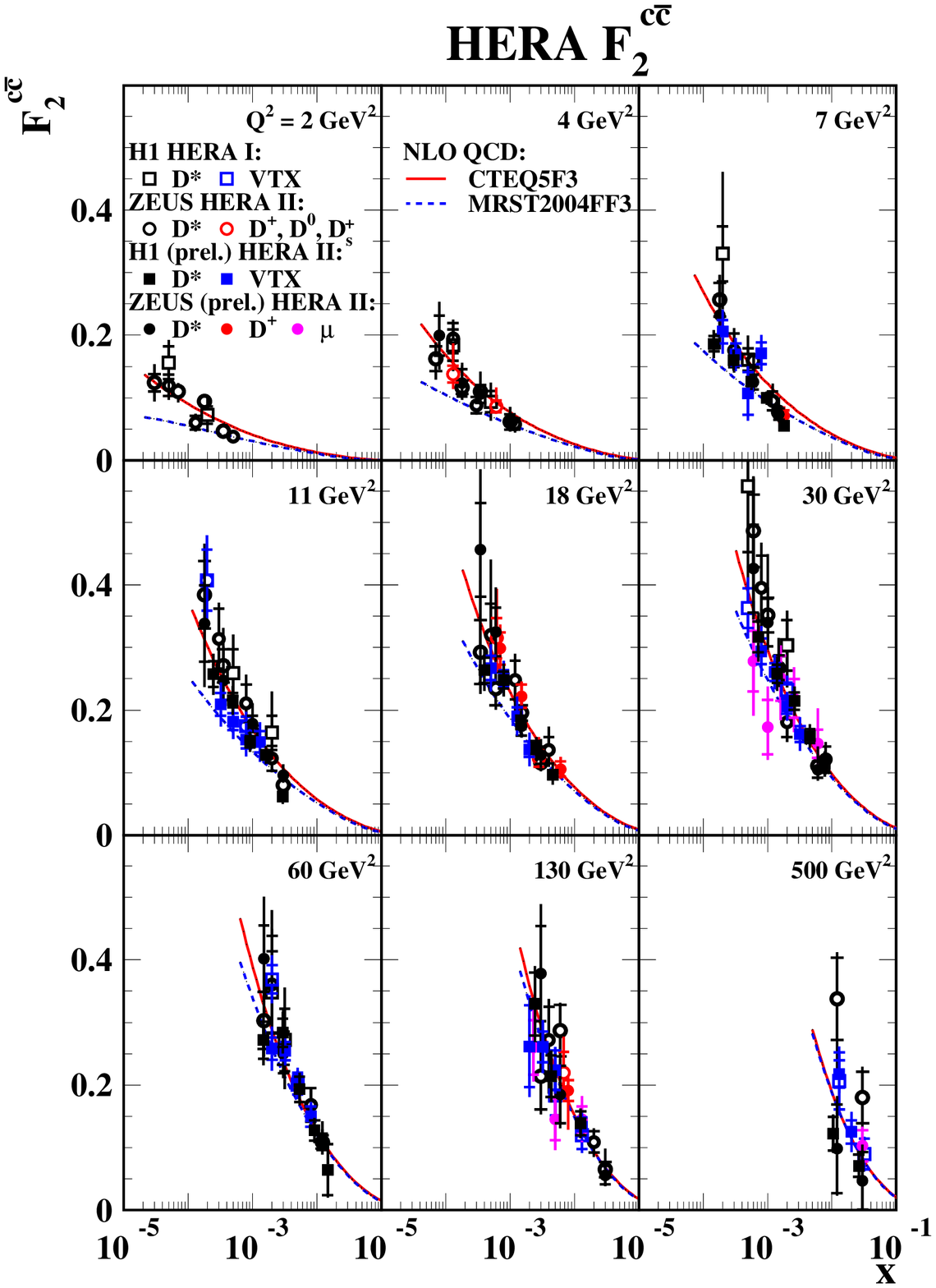}
\hfill \includegraphics[width=0.52\columnwidth]{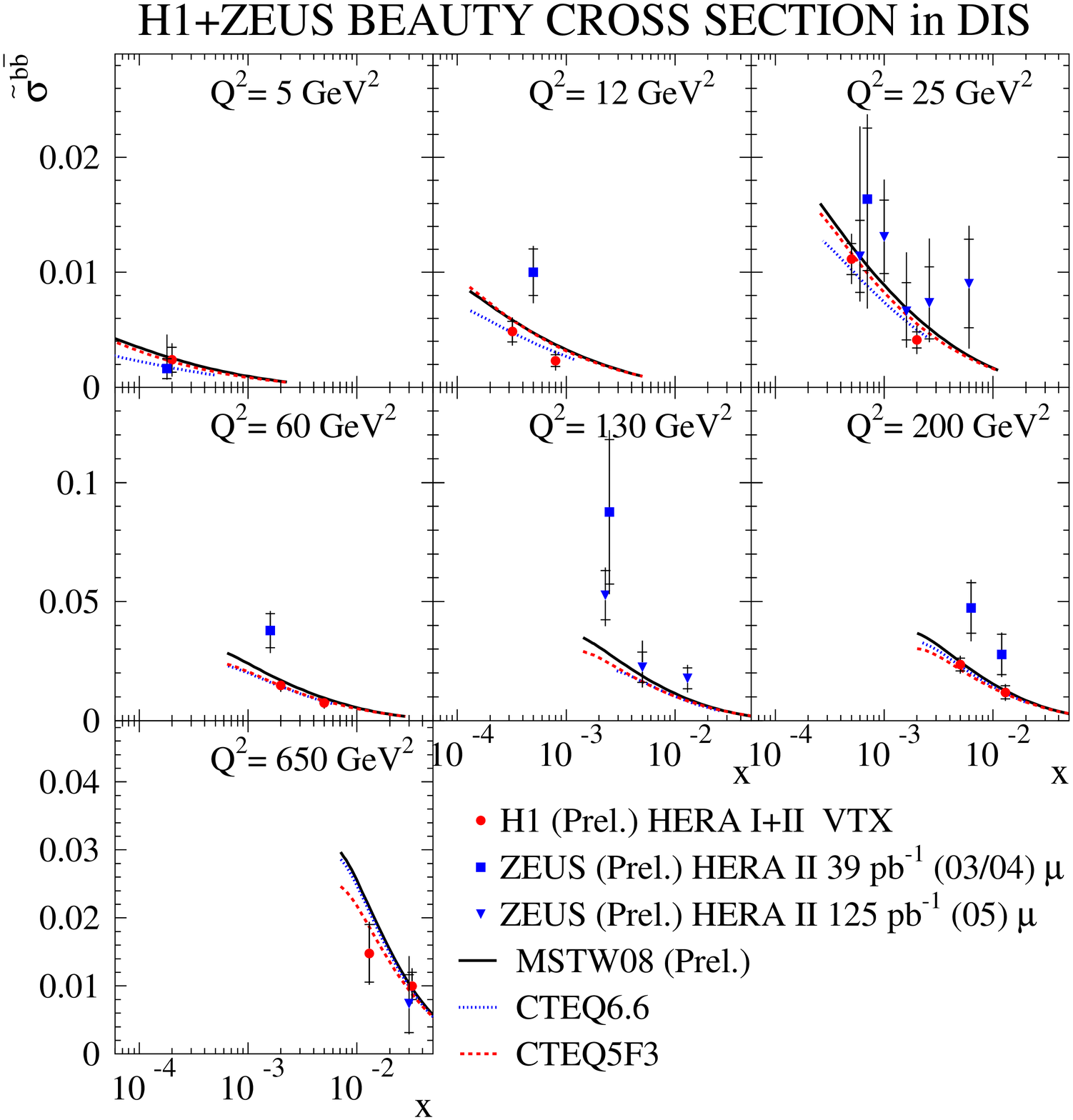}
\caption{Charm (left) and beauty (right) contribution to the proton structure
function $F_2$.\label{Fig:inclusive}} 
\end{figure*}

The charm and beauty contributions, $F_2^{c {\bar c}}$ and  $F_2^{b {\bar b}}$, 
to the inclusive proton structure function $F_2$ can be extracted from the
cross section measurements by extrapolating to the full phase space of the
studied final state. A different approach is employed in the inclusive lifetime
analysis, where all tracks with lifetime information from precise measurements
in the central silicon detector are used. The H1 Collaboration presented a
preliminary analysis~\cite{h1lifetime} based on this approach using a neural 
net to improve the separation power. A comparison of the different
measurements of $F_2^{c {\bar c}}$ and $F_2^{b {\bar b}}$ is shown in 
fig.~\ref{Fig:inclusive}. The experimental results agree well with 
each other. NLO QCD predictions with different proton parton distributions 
can describe the data and show the sensitivity of the data to the gluon 
density. 

\section{CHARM FRAGMENTATION}
One important ingredient to the extraction of the charm production cross section
and the charm contribution $F_2^{c{\bar c}}$ to the proton structure function
from measured $D$ meson cross sections is the fragmentation that describes the
transition of a charm quark into a charm hadron. The spectrum of the momentum 
fraction $z$ transferred from the heavy quark to the hadron influences strongly
the transverse momentum distribution of the heavy hadron and thus the
extrapolation from differential cross sections to $F_2^{c{\bar c}}$. Both 
ZEUS~\cite{zeusfrag} and H1~\cite{h1frag} have studied the fragmentation of
charm quarks. In the presence of a jet both experiments find results 
consistent with fragmentation functions determined from $e^+e^-$ data 
indicating that fragmentation is universal. For events with no jets,
corresponding to low photon gluon centre-of-mass energies, H1 finds  
a harder fragmentation which can be interpreted as an inadequacy of the QCD 
models to provide a consistent description of the full phase space down to the 
kinematic threshold.

\section{SUMMARY}
Several new measurements of the electroproduction of charm and beauty quarks at 
HERA from the H1 and ZEUS Collaborations were presented. In general the 
measurements are reasonably well described by the next-to-leading order 
perturbative QCD calculation HVQDIS. The charm and beauty contributions, 
$F_2^{c {\bar c}}$ and $F_2^{b {\bar b}}$, to the inclusive proton structure 
function $F_2$ extracted with different methods agree with each other and show 
sensitivity to the gluon density in the proton. Measurements of the charm
fragmentation function are an important input to the extrapolation of charm 
meson cross sections to $F_2^{c {\bar c}}$ and provide information on the 
universality of fragmentation.



%




\begin{acknowledgments}
This work was supported by the German Federal Ministry of Education and Research.
\end{acknowledgments}

\end{document}